\documentclass[referee]{aa} % for a referee version
\usepackage{bbding}
\usepackage{pifont}
\usepackage{amsmath}
\usepackage{graphicx}
\usepackage{stmaryrd}
\usepackage{txfonts}
\usepackage{url}
\usepackage{subfigure}
\usepackage{longtable}
\usepackage{lscape}
\usepackage{natbib}

\begin{document}
   \title{Multiwavelength study of the infrared dust bubble S51}

%   \subtitle{the infrared dust bubble S51}
    \authorrunning{C. P. Zhang \& J. J. Wang}
    \titlerunning{Bubble S51}

   \author{Chuan Peng Zhang
          \inst{1,2,3}
          \and
          Jun Jie Wang\inst{1,2}
          %\fnmsep
          %\thanks{Just to show the usage of the elements in the author field}
          }

   \institute{National Astronomical Observatories, Chinese Academy of Sciences,
             Beijing 100012, China\\
             \email{zcp0507@gmail.com}
         \and
             NAOC-TU Joint Center for Astrophysics, Lhasa 850000, China
         \and
             Graduate University of the Chinese Academy of Sciences, Beijing 100080, China
             %\email{c.ptolemy@hipparch.uheaven.space}
             %\thanks{The university of heaven temporarily does not accept e-mails}
             }

   \date{Received ---; accepted ---}

% \abstract{}{}{}{}{}
% 5 {} token are mandatory

  \abstract
   {}
{We investigate the environment of the infrared dust bubble S51 and
search for evidence of triggered star formation in its
surroundings.}
{We performed a multiwavelength study of the region around S51 with
data taken from large-scale surveys: 2MASS, GLIMPSE, MIPSGAL, IRAS,
and MALT90. We analyzed the spectral profile and the distribution of
the molecular gas ($^{13}$CO, C$^{18}$O, HCN, HNC, HCO$^{+}$,
C$_{2}$H, N$_{2}$H$^{+}$, and HC$_{3}$N), and dust in the
environment of S51. We used a mid-infrared emission three-color
image to explore the physical environment and GLIMPSE color-color
diagram [5.8]-[8.0] versus [3.6]-[4.5] to search for young stellar
objects and identify the ionizing star candidates.}
{From a comparison of the morphology of the molecular gas and the
Spitzer 8.0 $\mu$m emission, we conclude that the dust bubble is
interacting with CO at a kinematic distance of 3.4 kpc. The bubble
S51 structure, carried with shell and front side, is exhibited with
$^{13}$CO and C$^{18}$O emission. Both outflow and inflow may exist
in sources in the shell of bubble S51. We discover a small bubble
G332.646-0.606 ($R_{in}$ = 26$''$, $r_{in}$ = 15$''$, $R_{out}$ =
35$''$ and $r_{out}$ = 25$''$) located at the northwest border of
S51. A water maser, a methanol maser, and IRAS 16158-5055 are
located at the junction of the two bubbles. Several young stellar
objects are distributed along an arc-shaped structure near the S51
shell. They may represent a second generation of stars whose
formation was triggered by the bubble expanding into the molecular
gas.}
   {}


   \maketitle
%
%________________________________________________________________

\section{Introduction}    %% first-level sections will be auto-capitalized
\label{sect:intro}

\citet{chur2006,chur2007} detected and cataloged about 600
mid-infrared dust (MIR) bubbles between longitudes -60$^{\circ}$ and
+60$^{\circ}$. The bubbles have bright 8.0 $\mu$m shells that
enclose bright 24 $\mu$m interiors. The infrared (IR) dust bubbles
may be produced by exciting O- and/or B-type stars, which are
located inside the bubble. The ultraviolet (UV) radiation from
exciting stars may heat dust and ionize the gas to form an expanding
bubble shell \citep{wats2008}, which is known as the
''collect-and-collapse'' process. This process can trigger the
massive star formation near the shell clumps. These bubbles present
an important opportunity to study the interaction between HII
regions and molecular clouds.

We selected the IR dust bubble S51 from the catalog of
\citet{chur2006}. S51 is a complete (closed ring) IR dust bubble
centered on $l$=332.673, $b$=-0.618. It lies in the southern part of
RCW106 \citep{mook2004,wong2008,lo2009}. The radius of the shell is
1.48$'$ (1.47 pc), and the average thickness of the shell is 0.32$'$
(0.32 pc), while the eccentricity of the ellipse is 0.82. The
assumed distance is 3.4 kpc.

MALT90 is a pilot survey conducted with the Mopra telescope in
preparation for the Millimeter Astronomy Legacy Team Survey at 90
GHz \citep{fost2011}. One of the aims of this multimolecular line
mapping in this work is to examine how different molecules correlate
with each other and with bubble S51. The molecules selected include
$^{13}$CO, C$^{18}$O, HCN, HNC, HCO$^{+}$, C$_{2}$H, N$_{2}$H$^{+}$,
and HC$_{3}$N. $^{13}$CO has an optical depth of 1 to 12
\citep{wong2008}, but C$^{18}$O can provide optical depth and line
profile information ($n_{crit} = 3 \times 10^{3} cm ^{-3}$). MALT90
emission traces dense gas, and each molecular transition provides
slightly different information. For example, HCN ($n_{crit} = 2
\times 10^{5} cm ^{-3}$) traces high column density and is optically
thick; HNC ($n_{crit} = 3 \times 10^{5} cm ^{-3}$) is prevalent in
cold gas ($n_{crit} = 2 \times 10^{5} cm ^{-3}$); HCO$^{+}$
($n_{crit} = 4 \times 10^{5} cm ^{-3}$) often shows infall
signatures and outflow wings \citep{rawl2004,full2005};
N$_{2}$H$^{+}$ ($n_{crit} = 4 \times 10^{5} cm ^{-3}$) is more
resistant to freeze-out onto grains than the carbon-bearing species
\citep{berg2001}; HC$_{3}$N is a good tracer of hot core chemistry.

In this work, we report a multiwavelength study of the environment
surrounding the IR dust bubble S51. We aim to explore its
surrounding interstellar medium (ISM) and search for signatures of
star formation. We describe the data used in Sect.\ref{sect:data};
the results and discussion are presented in Sect.\ref{sect:results};
Sect.\ref{sect:summary} summarizes the results.

\section{Data}
\label{sect:data}

We analyzed IR and millimeter wavelength data extracted from the
following large-scale surveys: the Two Micron All Sky Survey
(2MASS)\footnote{2MASS is a joint project of the University of
Massachusetts and the Infrared Processing and Analysis
Center/California Institute of Technology, funded by the National
Aeronautics and Space Administration and the National Science
Foundation.} \citep{skru2006}, GLIMPSE \citep{benj2003,chur2009},
MIPSGAL \citep{care2009}, IRAS \citep{neug1984}, and MALT90
\citep{fost2011}.

GLIMPSE is a MIR survey of the inner Galaxy performed with the
Spitzer Space Telescope. We used the mosaicked images from GLIMPSE
and the GLIMPSE Point-Source Catalog (GPSC) in the Spitzer-IRAC
(3.6, 4.5, 5.8 and 8.0 $\mu$m). IRAC has an angular resolution
between $1.5''$ and $1.9''$ \citep{fazi2004,wern2004}. MIPSGAL is a
survey of the same region as GLIMPSE, using the MIPS instrument (24
and 70 $\mu$m) on Spitzer. The MIPSGAL resolution is 6$''$ at 24
$\mu$m.

The Mopra 22-m radio telescope has a full width at half maximum
(FWHM) beam size of $\sim33''$ at $\sim$110 GHz for $^{13}$CO $J = 1
- 0$ and C$^{18}$O $J = 1 - 0$ transitions. The correlator was
configured with 1024 channels over a 64-MHz bandwidth, which
provided a velocity resolution of $\thicksim$0.17 km s$^{-1}$ per
channel over a useable velocity bandwidth of $\sim$120 km s$^{-1}$
\citep{lo2009,bain2006,wong2008}. The observing was set up so that
the central channel corresponded to $\thicksim$50 km s$^{-1}$, at
which the velocity of the emission from the GMC complex is centered.

The on-the-fly (OTF) mapping mode of Mopra was used for MALT90. Maps
were made with the beam center running on a $3.4'\times3.4'$ grid.
The scan rate was $3.92''$ per second. The map was made with $12''$
spacing between adjacent rows, giving 17 rows per map. Because the
Mopra beam at 90 GHz is $38''$, this row spacing provides redundancy
in the map. The full 8 GHz bandwidth of the Mopra spectrometer
(MOPS) was split into 16 zoom bands of 138 MHz each, providing a
velocity resolution of $\thicksim$0.11 km s$^{-1}$ in each band. In
this work, six MALT90 pilot survey lines (HCN ($J = 1 - 0$), HNC ($J
= 1 - 0$), HCO$^{+}$ ($J = 1 - 0$), C$_{2}$H ($N = 1 - 0, J = 3/2 -
1/2, F = 2 - 1$), N$_{2}$H$^{+}$ ($J = 1 - 0$) and HC$_{3}$N ($J =
10 - 9$)) were used to trace the environment of bubble S51.

\begin{figure*}
\centering
\includegraphics[width=0.9\textwidth, angle=0]{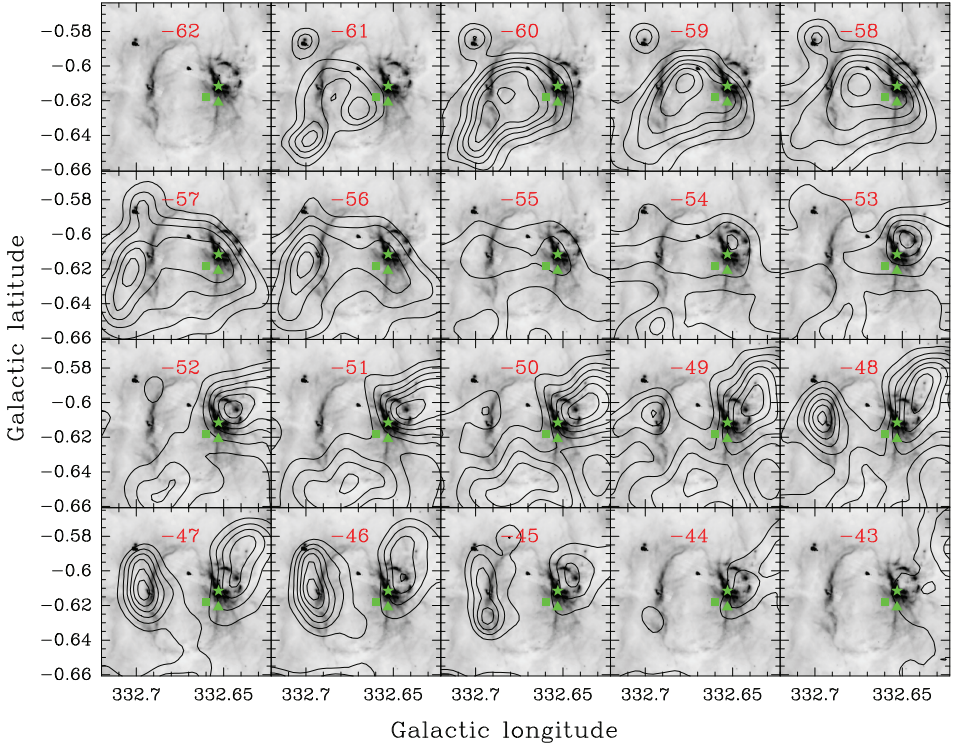}
\caption{Integrated velocity contours of the $^{13}$CO emission
every 1.0 km s$^{-1}$ superimposed on the GLIMPSE 8.0 $\mu$m
grayscale. The contours are at multiples of the 13\% level of each
of the emission peaks. The peak of each mosaic from -62 to -43 km
s$^{-1}$ is 0.718, 1.632, 3.584, 7.503, 9.191, 11.388, 14.901,
16.428, 10.915, 9.673, 12.091, 15.416, 16.916, 16.467, 14.378,
12.772, 6.043, 3.018, 1.535, and 2.631 K km s$^{-1}$, respectively.
The green symbols ''$\blacktriangle$'', ''$\blacksquare$'' and
''$\bigstar$'' indicate the positions of the water maser, the
methanol maser, and IRAS 16158-5055, respectively.}
\label{Fig:bubble_13co}
\end{figure*}

\begin{figure*}
\centering
\includegraphics[width=0.9\textwidth, angle=0]{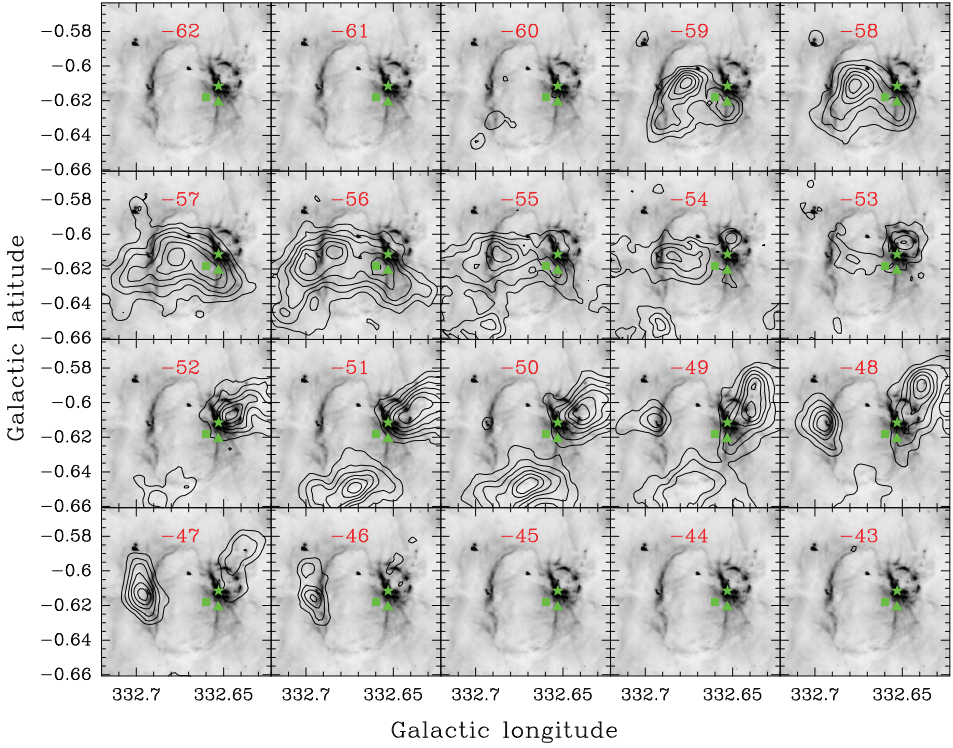}
\caption{Integrated velocity contours of the C$^{18}$O emission
every 1.0 km s$^{-1}$ superimposed on the GLIMPSE 8.0 $\mu$m
grayscale. The contours are at multiples of the 13\% level of each
of the emission peaks. The peak of each mosaic from -62 to -43 km
s$^{-1}$ is 0.518, 0.590, 0.870, 2.625, 4.342, 4.081, 3.663, 2.814,
1.627, 1.529, 2.671, 3.697, 4.914, 4.572, 4.514, 3.756, 1.524,
0.575, 0.558, and 0.685 K km s$^{-1}$, respectively. The green
symbols ''$\blacktriangle$'', ''$\blacksquare$'' and ''$\bigstar$''
indicate the positions of the water maser, the methanol maser, and
IRAS 16158-5055, respectively.} \label{Fig:bubble_c18o}
\end{figure*}

For the MOPRA observations, OFF positions were chosen at
$\pm$1$^{\circ}$ in Galactic latitude away from the plane (positive
offset for sources at positive Galactic latitude and vice versa),
and a single OFF position was observed for every two scan rows.
Pointing on SiO masers was performed every 1-1.5 hour, maintaining
pointing precision to better than about 10$''$. The brightness
temperature $T_{MB}$ is related to the antenna temperature
$T^{*}_{A}$ by $T_{MB} = T^{*}_{A}/\eta_{\nu}$, where $\eta_{\nu}$
is the frequency-dependent beam efficiency. According to
\citet{ladd2005} and \citet{lo2009}, the main beam efficiency at 86
GHz is $\eta_{86 GHz}$ = 0.49, and at 110 GHz is $\eta_{110 GHz}$ =
0.44. The results presented in this paper are in terms of main
bright temperature $T_{MB}(K)$. The $^{13}$CO and C$^{18}$O data
cubes were obtained from \citet{lo2009}. MALT90 data cubes were
downloaded from the online
archive\footnote{http://atoa.atnf.csiro.au/MALT90/}. The $^{13}$CO,
C$^{18}$O and MALT90 data cubes were processed with CLASS and GREG
in the GILDAS software
package\footnote{http://iram.fr/IRAMFR/GILDAS/}.

\section{Results and discussion}
\label{sect:results}

\begin{figure*}
\centering
\includegraphics[width=0.9\textwidth, angle=0]{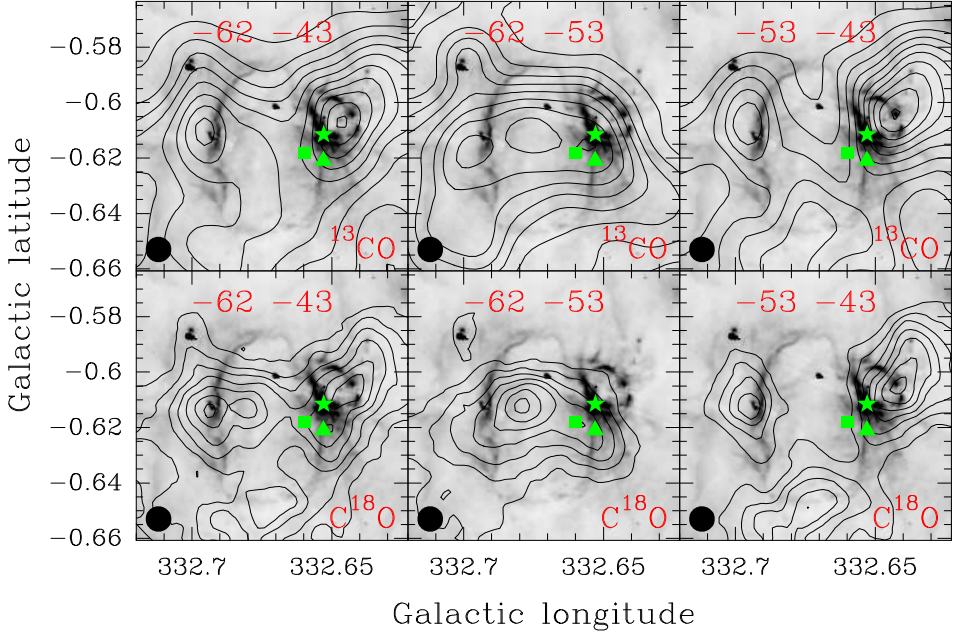}
\caption{Integrated velocity contours of the $^{13}$CO (Upper three
mosaics) and C$^{18}$O (Lower three mosaics) emission superimposed
on the GLIMPSE 8.0 $\mu$m grayscale. The contours are at multiples
of the 9\% and 12\% level of each of the $^{13}$CO and C$^{18}$O
emission peaks, respectively. The peaks of upper three mosaics from
left to right are 123.136, 59.055, and 93.380 K km s$^{-1}$; the
peaks of the lower three mosaics from left to right are 25.598,
18.456, and 20.458 K km s$^{-1}$, respectively. The integrated
velocity range is indicated in each panel. The green symbols
''$\blacktriangle$'', ''$\blacksquare$'' and ''$\bigstar$'' indicate
the positions of the water maser, the methanol maser, and IRAS
16158-5055, respectively.} \label{Fig:13co_c18o}
\end{figure*}

The clue of this section is as follows. Firstly, by analyzing the
channel maps (Figs.\ref{Fig:bubble_13co} and \ref{Fig:bubble_c18o})
of bubble S51 with $^{13}$CO and C$^{18}$O emission, we found that
bubble S51 has shell and front side characters strongly associated
with 8.0 $\mu$m emission. The integration intensity maps of [-62.0
-43.], [-62.0 -53.0] and [-53.0 -43.0] components are shown in
Fig.\ref{Fig:13co_c18o}. In addition, to examine the clumps of
bubble S51, we used MALT90 multimolecular line mapping
(Fig.\ref{Fig:mopra} in Appendix \ref{appendix}) to compare them
with each other and with IR emission distribution. The molecules
include HCN, HNC, HCO$^{+}$, C$_{2}$H, N$_{2}$H$^{+}$, and
HC$_{3}$N. And we also show the spectra of these molecules in
Fig.\ref{Fig:core_spectra}. These spectra indicate that there are
signs of inflow or outflow, whose velocity channels are shown in
Fig.\ref{Fig:outflow}. Furthermore, we show two three-color images
of MIR emission of IR dust bubble S51 with 3.6-4.5-8.0 $\mu$m and
4.5-8.0-24 $\mu$m in Fig.\ref{Fig:3_colors}, so that we can
understand the IR structure of bubble S51. Finally, we used GLIMPSE
color-color diagram [5.8]-[8.0] versus [3.6]-[4.5]
(Fig.\ref{Fig:allen}) and spectral energy distribution (SED) fitting
(Fig.\ref{Fig:sed}) to search for young stars and identify exciting
star candidates, whose distribution is shown in Fig.\ref{Fig:stars}.

\subsection{The shell and front side of bubble S51}

We inspected the molecular gas around S51 from the Mopra $^{13}$CO
and C$^{18}$O emission in the whole velocity range and found two
independent velocity components ([-62.0 -53.0] and [-53.0 -43.0] km
s$^{-1}$) along the line of sight. Figs.\ref{Fig:bubble_13co} and
\ref{Fig:bubble_c18o} show the integrated velocity maps of the
$^{13}$CO and C$^{18}$O emission every 1.0 km s$^{-1}$ between -62.5
and -42.5 km s$^{-1}$. The 8.0 $\mu$m grayscale shows the position
and size of S51.

Comparing the molecular and MIR morphologies, there are many
interesting places among the [-62.0 -53.0], [-53.0 -43.0] km
s$^{-1}$ components and 8.0 $\mu$m emission, shown in
Figs.\ref{Fig:bubble_13co}, \ref{Fig:bubble_c18o} and
\ref{Fig:13co_c18o}. The morphology and peaks of [-53.0 -43.0] km
s$^{-1}$ component clearly correlate with 8.0 $\mu$m emission.
Accordingly, the [-53.0 -43.0] km s$^{-1}$ component may be the
shell of bubble S51. On the other hand, looking at the -58.0 km
s$^{-1}$ panel in Fig.\ref{Fig:bubble_c18o}, we find that the
contours fit quite well within the 8.0 $\mu$m emission ring. The
integration velocity map of the [-62.0 -53.0] km s$^{-1}$ component
in Fig.\ref{Fig:13co_c18o} also shows that there is a good
correlation between C$^{18}$O contours and 8.0 $\mu$m emission. We
suggest that this cloud could be blueshifted gas associated with the
front side of bubble S51. Since the -48.0 km s$^{-1}$ cloud is well
associated with bubble shell, this bubble would be expanding at
roughly 10.0 km s$^{-1}$ along the line of sight. However, there is
no redshifted counterpart.

\begin{figure}
\centering
\includegraphics[width=0.7\textwidth, angle=0]{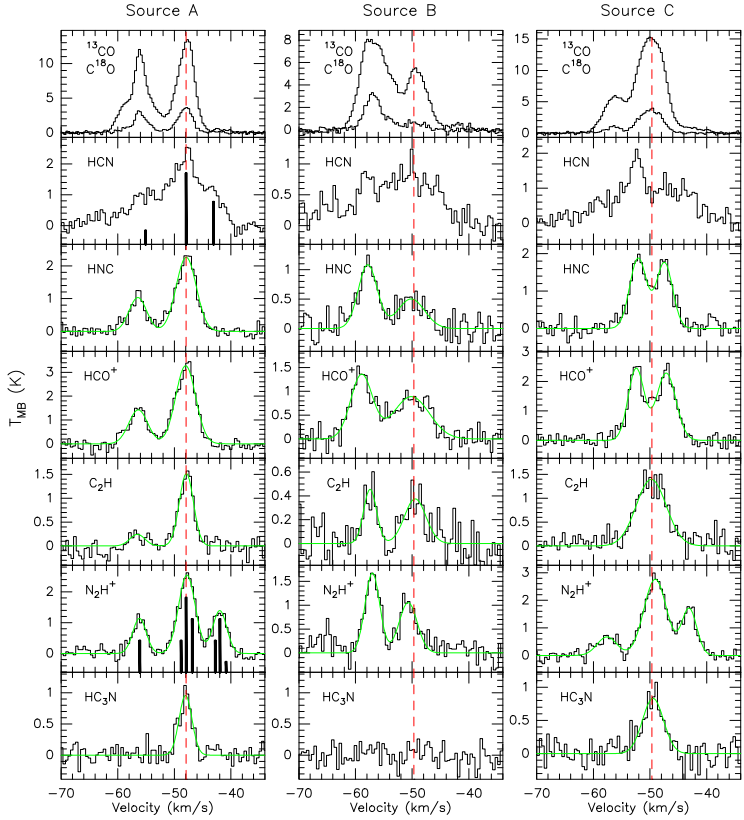}
\caption{MALT90 spectra of sources A, B and C, whose positions are
represented in Fig.\ref{Fig:stars} with ''A'' ($l$=332.693,
$b$=-0.610),''B'' ($l$=332.677, $b$=-0.619) and ''C'' ($l$=332.645,
$b$=-0.608 ), respectively. The red dashed lines indicate the peak
velocities of the C$^{18}$O emission. The green curves are Gauss-fit
lines. HCN and N$_{2}$H$^{+}$ have hyperfine splitting, shown as a
black bar.} \label{Fig:core_spectra}
\end{figure}

From the $^{13}$CO and C$^{18}$O spectra of
Fig.\ref{Fig:core_spectra}, we can also see that there are two
velocity components at the locations of sources A, B and C (see
Fig.\ref{Fig:stars}), where source A is the core of eastern lobe of
S51, source B lies at the front side or inside S51, and source C is
the core of western lobe of S51. It is obvious that the spectral
intensity of the [-62.0 -53.0] km s$^{-1}$ component at the front
side of bubble is stronger than that of [-53.0 -43.0] km s$^{-1}$.
At the shell of bubble S51, however, it is just the reverse. This
case is also evidence that bubble S51 has a shell and front side
structure.

For the [-53.0 -43.0] km s$^{-1}$ component, the optical thin
C$^{18}$O emission contours and spectra (Fig.\ref{Fig:bubble_c18o}
and \ref{Fig:core_spectra}) show that the systematic velocity is
$\thicksim$-50.0 km s$^{-1}$. Assuming the galactic rotation model
of \citet{fich1989} (with $R_{\odot}$ = 8.5 kpc and $\Theta_{\odot}$
= 220 km s$^{-1}$), we obtain kinematic distances of either 3.4 or
11.7 kpc. \citet{lock1979} suggested that RCW106 lies the near
kinematic distance after examining H$_{2}$CO absorption spectra seen
against the HII region continuum. Therefore, we adopted the near
kinematic distance of 3.4 kpc in this work.

\subsection{Opacity, excitation temperature, and column density}

We used the approach of \citet{wong2008} to derive the opacity,
excitation temperature, and column density of $^{13}$CO and
C$^{18}$O within the box shown in Fig. \ref{Fig:mopra} of Appendix
\ref{appendix}. The size of the box is $4.3'\times4.3'$ (where
$4.3'$ is the diameter of S51). The velocity range included is just
$V$ = $\thicksim$-53.0 to $\thicksim$-43.0 km s$^{-1}$ of bubble
shell, and the box is centered on $l$=332.673, $b$=-0.618.

The relation between opacities and the ratio of $^{13}$CO to
C$^{18}$O main-beam brightness temperature \citep{myer1983} is
\begin{equation}
    \label{eq:13co2c18o}
    \frac{T_{MB}(^{13}CO)}{T_{MB}(C^{18}O)}=\frac{1-exp(-\tau_{13})}{1-exp(-\tau_{18})}=\frac{1-exp(-7.4\tau_{18})}{1-exp(-\tau_{18})}.
\end{equation}
Equation (\ref{eq:13co2c18o}) assumes $\tau_{13} = 7.4 \tau_{18}$
\citep{wils1994,wong2008}. Furthermore, the excitation temperature
$T_{ex}$ is derived from the equation of radiative transfer:
\begin{equation} \label{eq:tmb_jt}
\left\{ \begin{aligned}
         T_{MB}=f[J(T_{ex})-J(T_{bg})][1-exp(-\tau)] \\
         J(T)=T_{0}/[exp(T_{0}/T)-1]
\end{aligned}, \right.
\end{equation}
where $f$ is the beam filling factor, $T_{bg}$ = 2.7 K and
$T_{0}=h\nu/k$ = 5.29 K for the 1-0 transition of $^{13}$CO
\citep{wong2008}. And then, assuming $^{13}$CO is optical thin, we
obtain the molecular $^{13}$CO column density $N(^{13}$CO) from the
relation \citep{bour1997}
\begin{small}
\begin{equation}
    \label{eq:tmb}
    N(^{13}CO)_{thin}=\frac{T_{ex}+0.88}{1-exp(-5.29/T_{ex})}\cdot\frac{2.42\times10^{14}}{J(T_{ex})-J(T_{bg})}\int
    T_{MB}(^{13}CO)dv.
\end{equation}
\end{small}
Finally, assuming [H$_{2}$/$^{13}$CO] abundance ratio is
$7\times10^{5}$ \citep{frer1982}, the molecular hydrogen column
$N(H_{2})$ was calculated. The molecular cloud mass was estimated at
$M_{shell} \sim 1.3\times10^{4} M_{\odot}$ for the bubble shell at a
distance of $D_{S51} \sim 3.4$ kpc.

\subsection{Clump analysis}

\begin{figure}
\centering
\includegraphics[width=0.7\textwidth, angle=0]{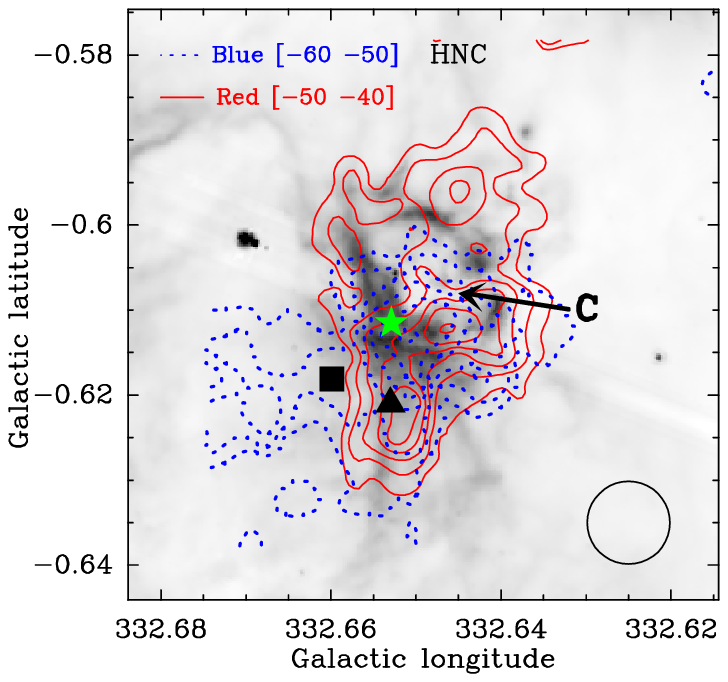}
\includegraphics[width=0.7\textwidth, angle=0]{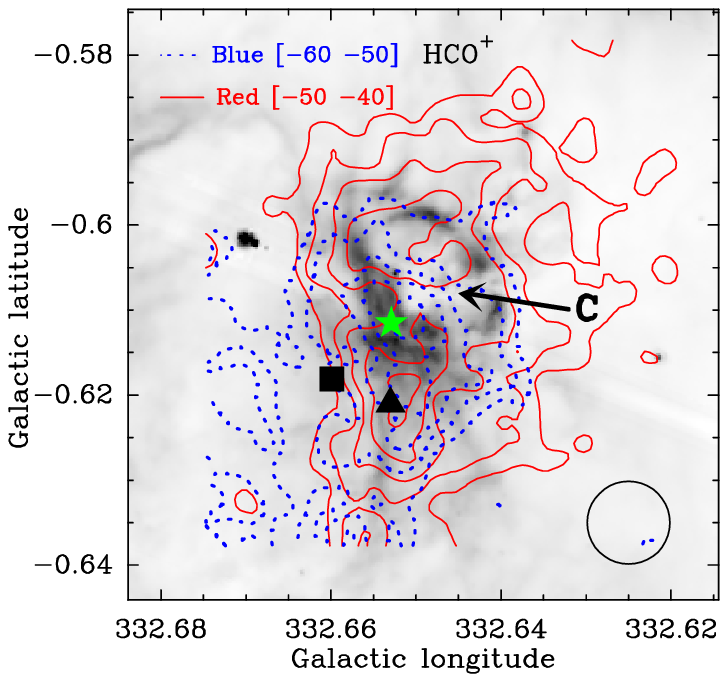}
\caption{Integrated velocity contours of HNC (top panel) and HCO$^+$
(bottom panel) emission superimposed on the GLIMPSE 8.0 $\mu$m
grayscale. The contours are at multiples of 11\%, 11\%, 12\%, and
12\% level of the blue lobe of HNC, the red lobe of HNC, the blue
lobe of HCO$^+$, and the red lobe of HCO$^+$ emission peaks, which
are 9.539, 8.618, 12.861, and 10.800 K km s$^{-1}$, respectively.
Source C is indicated with ''C'' in the image. The symbols
''$\blacktriangle$'', ''$\blacksquare$'' and ''$\bigstar$'' indicate
the positions of the water maser, the methanol maser, and IRAS
16158-5055, respectively.} \label{Fig:outflow}
\end{figure}

To examine the clumps of bubble S51, we used multimolecular line
mappings to compare them with each other and with the IR emission
distribution.

\begin{figure}
\centering
\includegraphics[width=0.65\textwidth, angle=0]{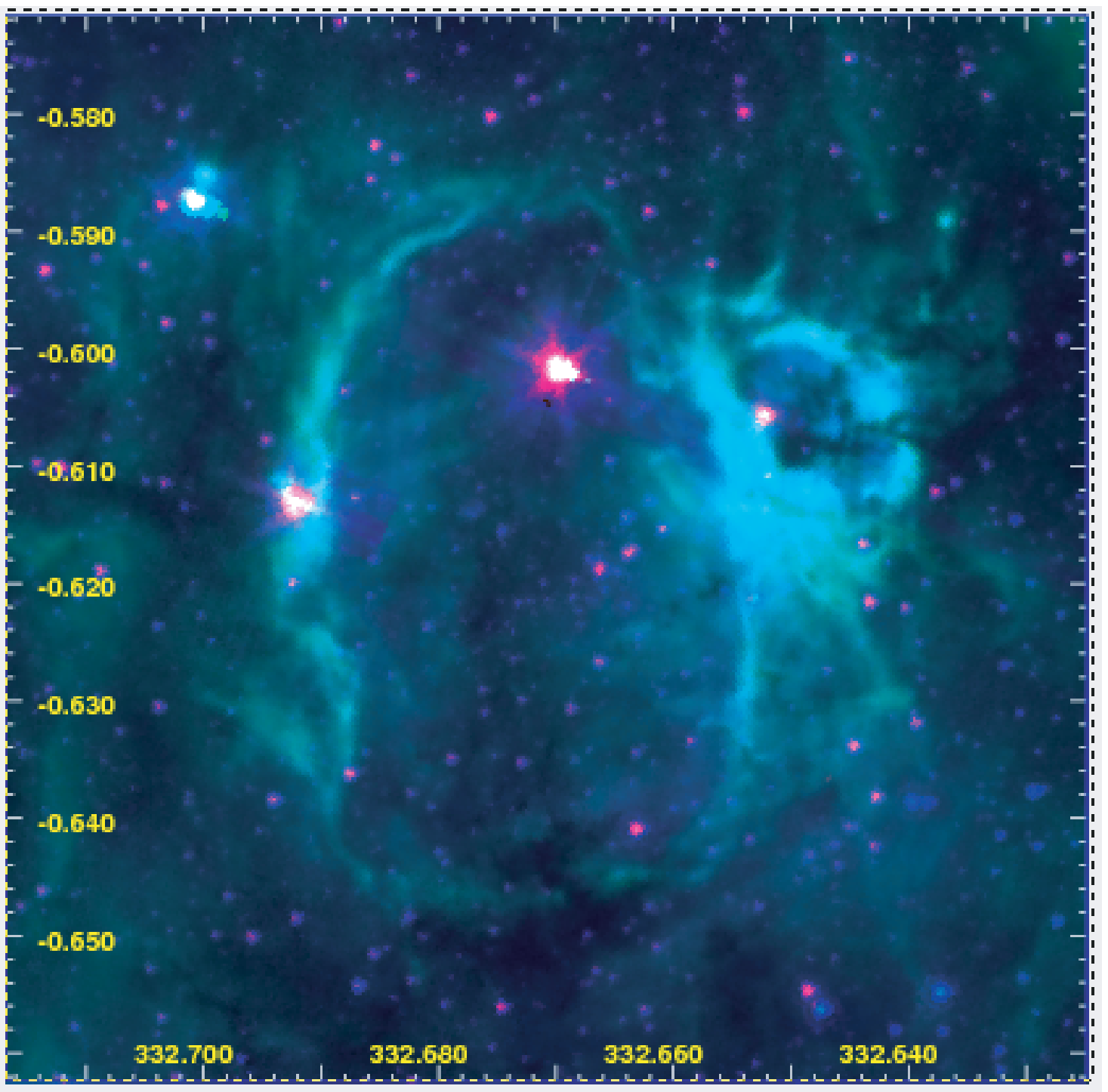}
\includegraphics[width=0.65\textwidth, angle=0]{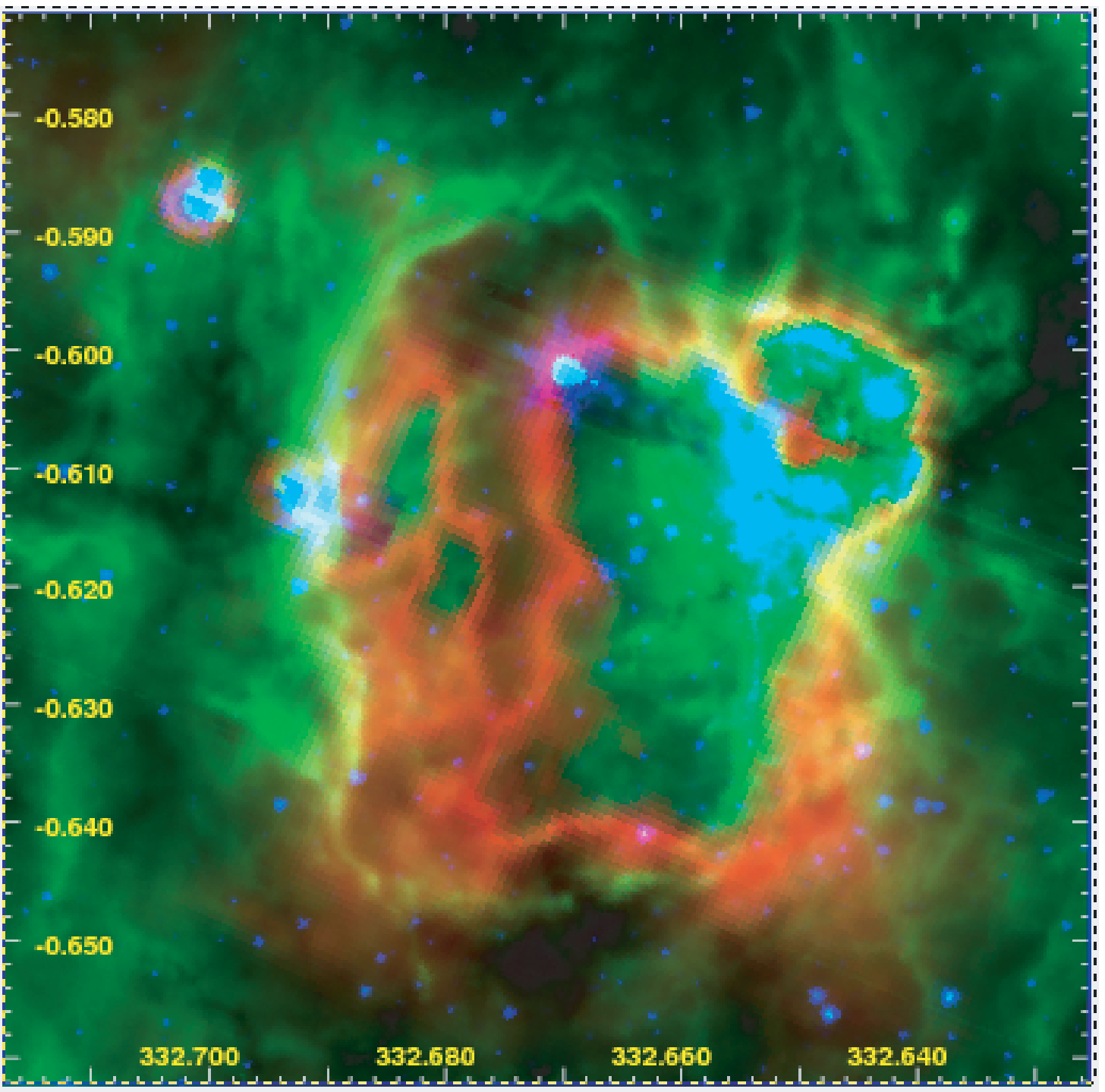}
\caption{MIR emission of IR dust bubble S51. $Top$: Spitzer-IRAC
three-color image (3.6 $\mu$m = red, 4.5 $\mu$m = blue and 8.0
$\mu$m = green); $Bottom$: Spitzer-IRAC and Spitzer-MIPSGAL
three-color image (4.5 $\mu$m = blue, 8.0 $\mu$m = green and 24
$\mu$m = red). Note that the 24 $\mu$m emission is saturated at the
center of image.} \label{Fig:3_colors}
\end{figure}

In the box shown in Fig. \ref{Fig:mopra}, the maximum and average
column densities N($^{13}$CO) are 9.72$\times$10$^{17}$ and
5.98$\times$10$^{16}$ cm$^{-2}$, respectively. This figure also
shows the main beam brightness temperature $T_{MB}$, the brightness
temperature ratio $T_{MB}$($^{13}$CO)/$T_{MB}$(C$^{18}$O), the
optical depth $\tau$($^{13}$CO), and the excitation temperature
$T_{ex}$($^{13}$CO). The distribution of these parameters (shown in
color) are compared with contours showing the integrated intensity
of different molecular transitions. The average
$T_{MB}$($^{13}$CO)/$T_{MB}$(C$^{18}$O), $\tau$($^{13}$CO) and
$T_{ex}$($^{13}$CO) are 4.49, 1.76 and 20.91, respectively. The peak
positions of $^{13}$CO, C$^{18}$O and 8.0 $\mu$m emission are
spatially coincident, and are situated at the east and west edge of
the shell. The molecular clouds are extended around the 8.0 $\mu$m
emission peaks. At the south border of bubble S51, we can also see
the shell of the filament structure from 8.0 $\mu$m emission in
Fig.\ref{Fig:3_colors}. The $^{13}$CO and C$^{18}$O emission
contours are distributed along the shell of the filament structure
in Fig.\ref{Fig:13co_c18o}. We also see from Fig.\ref{Fig:mopra}
that the distribution of $T_{MB}$($^{13}$CO)/$T_{MB}$(C$^{18}$O),
$\tau$($^{18}$CO), $T_{ex}$($^{13}$CO) and N($^{13}$CO) shows an
arc-like morphology around the north edge of S51. However, it should
be noted that these derived properties are not independent of each
other.

HCN, HNC, HCO$^{+}$, C$_{2}$H, and N$_{2}$H$^{+}$ trace the position
of highest density, while HC$_{3}$N traces the location of the hot
core \citep{fost2011}. We chose three sources (A, B and C) to
analyze in Fig.\ref{Fig:stars}; their spectra are shown in
Fig.\ref{Fig:core_spectra}. Fig.\ref{Fig:mopra} shows that a
potentially dense and hot core is mainly located in the eastern and
western shells. The spectra shown in Fig.\ref{Fig:core_spectra} for
the [-53.0 -43.0] km s$^{-1}$ component demonstrate that source B
inside S51 has a lower brightness temperature than sources A and C
in the S51 shell. At source B, however, the MALT90 spectra of the
[-53.0 -43.0] km s$^{-1}$ component show the blueshifted velocity
[-62.0 -53.0] km s$^{-1}$, which is evidence that this is the front
side of bubble S51. This suggests that the extended cloud of the
[-53.0 -43.0] km s$^{-1}$ component is scarce inside bubble S51, and
the front side of bubble S51 has a rich abundance. However, the back
side of bubble S51 is not shown in the MALT90 spectra. They also
show that bubble S51 might be expanding at 10 km s$^{-1}$ compared
to the bubble shell along the line of sight. In addition, we note
that IRAS 16158-5055 is closer to the peaks of N$_{2}$H$^{+}$ and
HC$_{3}$N than the water maser and the methanol maser.

\subsection{Inflow or outflow character}

As shown in Fig.\ref{Fig:outflow}, there is much trace of active
star formation located in the shell of S51. The western shell of S51
contains a UCHII region, a water maser ($l$=332.653, $b$=-0.621), a
methanol maser ($l$=332.660, $b$=-0.618), and IRAS 16158-5055
($l$=332.653, $b$=-0.612) \citep{wals1997,bree2007}. The water maser
spectrum has three peaks, with fluxes of $\sim$1 Jy for the -58.5 km
s$^{-1}$ component, $\sim$6 Jy for -47.5 km s$^{-1}$, and $\sim$30
Jy for -45.2 km s$^{-1}$, respectively \citep{wals1997}. The
methanol maser spectrum has a velocity range of -52 $\sim$ -50 km
s$^{-1}$, and the peak flux is 7.1 Jy \citep{bree2007}. This means
that the [-53.0 -43.0] km s$^{-1}$ component of the S51 and not the
[-62.0 -53.0] km s$^{-1}$ component of background is strongly
correlated with the methanol and water masers.

\begin{figure}
\centering
\includegraphics[width=0.6\textwidth, angle=0]{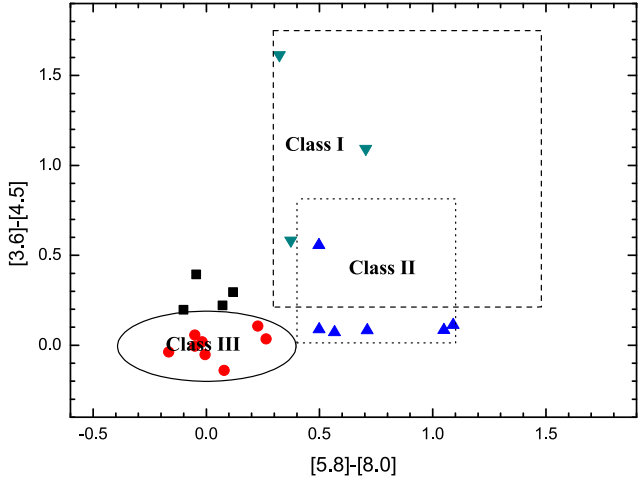}
\caption{GLIMPSE color-color diagram [5.8]-[8.0] versus [3.6]-[4.5]
for sources in and around S51. The symbols ''$\blacktriangledown$'',
''$\blacktriangle$'', ''$\medbullet$'' and ''$\blacksquare$''
indicate the different stars of class I, II, III and others,
respectively. The classification of class I, II and III indicates
the stellar evolutionary stage as defined by \citet{alle2004}.}
\label{Fig:allen}
\end{figure}

On the west side of S51, the position of source C is marked in
Fig.\ref{Fig:outflow}. Fig.\ref{Fig:core_spectra} shows the
$^{13}$CO, C$^{18}$O and MALT90 spectra of source C. We marked the
splittings of the hyperfine structure of HCN and N$_{2}$H$^{+}$ with
several black bars. The [-62.0 -53.0] km s$^{-1}$ component is very
faint in $^{13}$CO, C$^{18}$O emission, as well as HCN, HNC,
HCO$^+$, C$_{2}$H, and HC$_3$N. Consequently the [-62.0 -53.0] km
s$^{-1}$ component seems unrelated to the [-53.0 -43.0] km s$^{-1}$
component in this cluster. Fig.\ref{Fig:core_spectra} also shows
that the optically thick spectra of HCN, HNC and HCO$^+$ lines have
absorption dips at the peak velocity (-49.5 km s$^{-1}$) traced by
C$^{18}$O, C$_{2}$H, N$_{2}$H$^{+}$ and HC$_3$N. The blue profile is
stronger than the red profile shown from the HCN spectrum. Such a
profile can be produced by inflow with cold gas wings and a hot
central core \citep{mard1997}. Moreover, HNC and HCO$^+$ have broad
line wings corresponding to N$_{2}$H$^{+}$ and HC$_3$N, therefore
source C could also be produced by outflow. Fig.\ref{Fig:outflow}
shows the emission in two different velocity channels. The
velocities of the water maser and the methanol maser are similar to
the more negative velocity gas. We speculate that since this region
is located at the junction of bubble S51 and bubble G332.646-0.606
(referred to Sect.\ref{sec:irstru}), it is plausible that the
interactions between the two bubbles or between the HII region and
the molecular cloud have triggered this phenomenon.

\subsection{IR structure of S51} \label{sec:irstru}

Fig.\ref{Fig:3_colors} shows two Spitzer three-color images of S51.
Both figures clearly show the PDR visible in 8.0 $\mu$m (in green)
emission, which originates mainly in the PAHs. Since these large
molecules are destroyed inside the ionized region, the PAH emission
delineates the HII region boundaries, and the molecular clouds are
excited in the PDR by the radiation leaking from the HII region
\citep{petr2010,poma2009}. The 24 $\mu$m emission (bottom panel in
Fig.\ref{Fig:3_colors}) appears to be inside bubble S51, and
corresponds to hot dust. It is likely that O- and/or early B-type
stars produced the bubble shell of this HII region, with hot dust
located inside the bubble. The 3.6 $\mu$m emission (in red, top
panel in Fig.\ref{Fig:3_colors}) shows the positions of the
brightest stars.

\begin{figure}
\centering
\includegraphics[width=0.7\textwidth, angle=0]{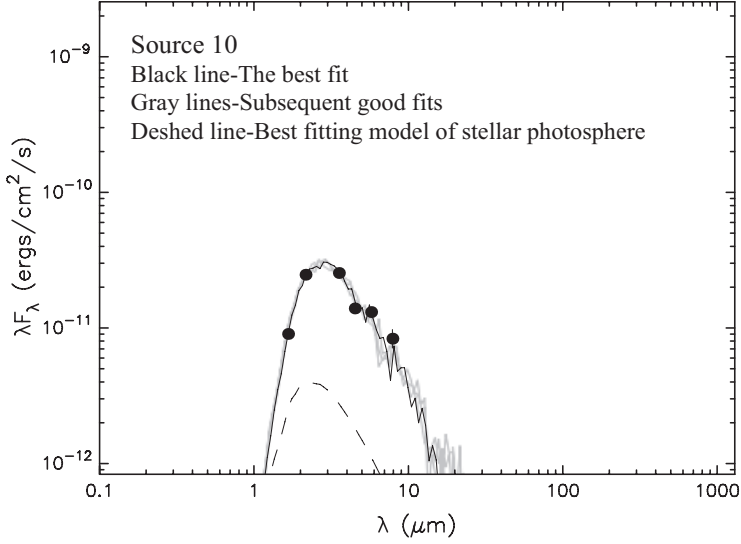}
\caption{SED fits of exciting star source 10 ($l$=332.6664,
$b$=-0.6266). The flux density of fitted data are the 2Mass H-band
5.01(0.22), the K-band 17.81(0.44), IRAC 3.6 $\mu$m 30.16(1.26), 4.5
$\mu$m 20.93(1.48), 5.8 $\mu$m 25.12(1.53), and 8.0 $\mu$m
22.15(3.46) mJy, respectively.} \label{Fig:sed}
\end{figure}

We also note a small bubble G332.646-0.606, which is centered on
$l$=332.646, $b$=-0.606 and located at the northwest edge of S51.
The semimajor ($R_{in}$) and semiminor ($r_{in}$) axes of the inner
ellipse are 26 and 15 arcsec, respectively; the semimajor
($R_{out}$) and semiminor ($r_{out}$) axes of the outer ellipse are
35 and 25 arcsec, respectively. From 8.0 $\mu$m emission of bubble
G332.646-0.606, the elliptical morphologic PDR shows a sketch of the
bubble. Note that the 24 $\mu$m emission is saturated inside the
small bubble G332.646-0.606. At the junction between bubble S51 and
G332.646-0.606, the IR emission is brighter than the rest of the
shell; this is also true for the $^{13}$CO, C$^{18}$O and MALT90
emission lines. IRAS 16158-5055 is also located at this position. It
seems likely that this junction region is the birth space of several
YSOs. In addition, the molecular gas of the bubble shell exhibits
several clumps along the PDR (Fig.\ref{Fig:3_colors}). The
distribution and morphology of this material suggests that a
collect-and-collapse process may be occurring.

\subsection{Search for young stars and identifying exciting stars}

The distribution of the IR point sources in the surroundings of S51
provides some signs of star formation. Fig.\ref{Fig:allen} shows the
[5.8]-[8.0] versus [3.6]-[4.5] color-color diagram for the sources
extracted from the GLIMPSE Point Source Catalog in the Spitzer-IRAC
bands in and around S51. We only considered sources with detection
in the four Spitzer-IRAC bands. In Fig.\ref{Fig:allen}, class I
stars are protostars with circumstellar envelopes; class II are
disk-dominated objects; and class III are main sequence and giant
stars. The classification indicates the stellar evolutionary stage
as defined by \citet{alle2004}.

\begin{figure}
\centering
\includegraphics[width=0.8\textwidth, angle=0]{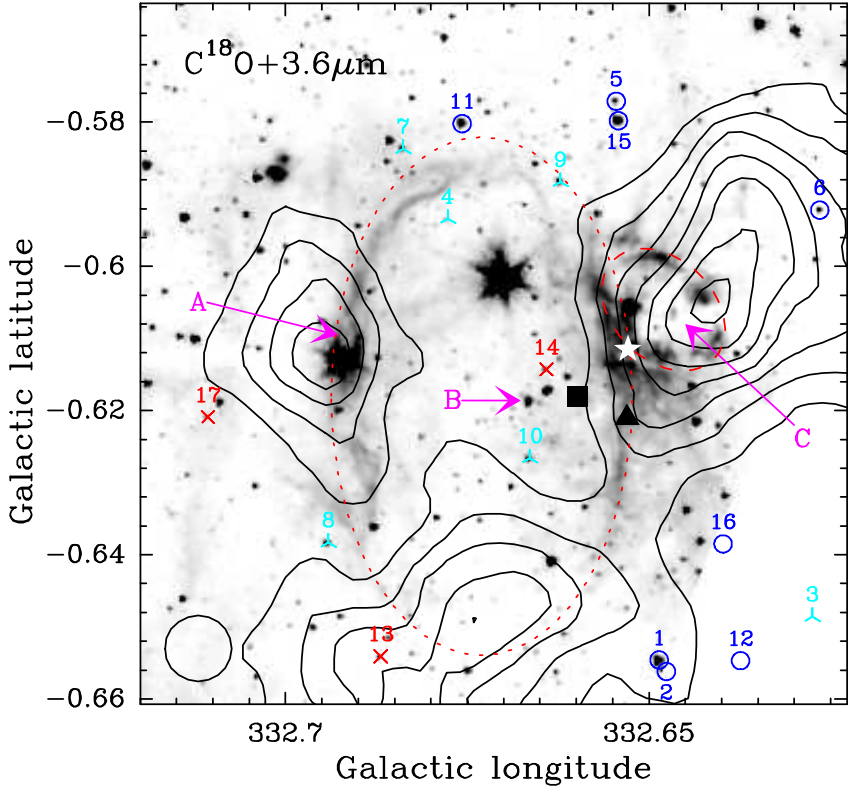}
\caption{Integrated velocity contours of the $^{18}$CO emission
superimposed on the GLIMPSE 3.6 $\mu$m grayscale. The contours are
at multiples of 12\% level of the C$^{18}$O emission peaks 93.380 K
km s$^{-1}$, and the integrated velocity range is from -53.0 to
-43.0 km s$^{-1}$. Two dashed ellipses show the PAH morphology of
the two bubbles. Letters ''A'' ($l$=332.693, $b$=-0.610),''B''
($l$=332.677, $b$=-0.619) and ''C'' ($l$=332.645, $b$=-0.608)
indicate the chosen positions of three sources. The class I, II and
III stars are indicated with symbols ''$\times$'', ''$\Yup$'' and
''$\medcirc$''. The symbols ''$\blacktriangle$'', ''$\blacksquare$''
and ''$\bigstar$'' indicate the positions of the water maser, the
methanol maser, and IRAS point source, respectively.}
\label{Fig:stars}
\end{figure}

Class I and class II stars are chosen to be the YSO candidates. The
YSO candidates are drawn with symbols ''$\times$'' and ''$\Yup$'' in
Fig.\ref{Fig:stars}, among which sources 4, 7, 8, 9, 13 and 17 near
the shell of bubble are made up of an arc-shaped distribution. This
may be caused by interaction between the HII region and molecular
clouds, and be the production of a collect-and-collapse process.

We also tried to search for and identify exciting stars of S51.
However, only sources 10 and 14 are located within the bubble.
Source 14 belongs to a class I star, therefore we did not consider
it to be an exciting star. Source 10 is a class II star, therefore
we identified it in more detail with an SED fitting, using the tool
developed by \citet{robi2007}, which is available
online\footnote{http://caravan.astro.wisc.edu/protostars/}. The good
fitting models were selected according to the condition
$\chi^{2}-\chi^{2}_{min} < 3$, where $\chi^{2}_{min}$ is the minimum
value of $\chi^{2}$ among all models.

In Fig.\ref{Fig:sed}, we show the results for the SED fitting of our
source 10 to the fluxes obtained from the GLIMPSE Point Source
Catalog \citep{hora2008} and the 2MASS All-Sky Point Source Catalog
\citep{skru2006}. We fit these sources allowing the extinction to
range from 0 to 30 mag and the distance to range from 3.0 to 4.0
kpc. The SED output shows that source 10 in class II is
$\dot{M}_{env}/M_{\star} < 10^{-6} yr^{-1}$ and $M_{disk}/M_{\star}
< 10^{-6}$; the temperature of the center star is 31000 K, in
agreement with the effective temperature expected for an O-type star
\citep{scha1997}. Therefore, source 10 may be a candidate of the
ionization of S51. Stellar winds emit from exciting stars. In
addition, the presence of a central cavity in the distribution of
the ionized gas may be a signature of the activities of stellar
winds. The location of source 10 inside this cavity provides
additional support for the assumption that this star is the most
likely candidate to have created S51. Spectroscopic observation is
needed to confirm this.

\section{Summary}
\label{sect:summary}

We have investigated the environment of the IR dust bubble S51 with
several spectra ($^{13}$CO, C$^{18}$O, HCN, HNC, HCO$^{+}$,
C$_{2}$H, N$_{2}$H$^{+}$, and HC$_{3}$N) and IR emission. The main
results can be summarized as follows.

(1) We have distinguished two independent velocity components
associated with S51 along the line of sight. One component belongs
to the shell of bubble S51, and another may be the front side of
bubble S51.

(2) We suggest that source C ($l$=332.645, $b$=-0.608) shows
evidence of either outflow or inflow, located on the northwest
border of S51.

(3) Next to the west border of bubble S51, we found a small bubble
G332.646-0.606, whose dimensions are $R_{in}$ = 26$''$, $r_{in}$  =
15$''$, $R_{out}$ = 35$''$ and $r_{out}$ = 25$''$.

(4) MALT90 emission contours appear to be correlated with the S51
shell. The water maser, the methanol maser, and IRAS 16158-5055 are
located in the shell. The YSO distribution is also correlated with
$^{13}$CO, C$^{18}$O and 8.0 $\mu$m emission.

\begin{acknowledgements}
We wish to thank the anonymous referee for comments and suggestions,
which improved the clarity of the paper. Particularly, we thank
Nadia Lo for providing $^{13}$CO and C$^{18}$O data. This work was
supported by the Young Researcher Grant of National Astronomical
Observatories, Chinese Academy of Sciences No.O835032002.
\end{acknowledgements}

\bibliographystyle{aa}
\bibliography{references}

\Online \onecolumn
\begin{appendix}
\section{Molecular line maps of bubble S51}\label{appendix}
\begin{figure*}[hp]
\centering
\includegraphics[width=0.95\textwidth, angle=0]{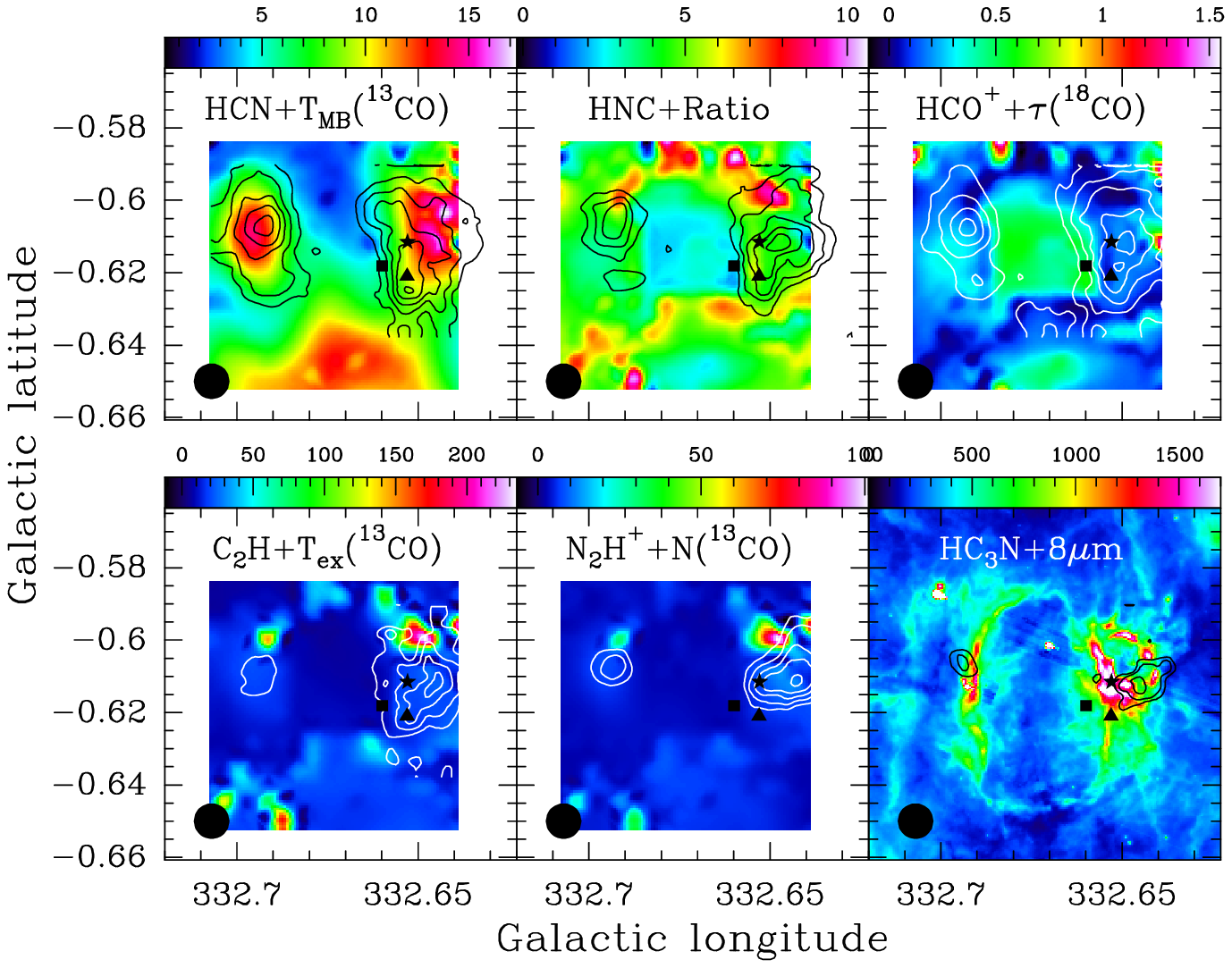}
\caption{Integrated velocity contours of MALT90 emissions (HCN, HNC,
HCO$^{+}$, C$_{2}$H, N$_{2}$H$^{+}$, and HC$_{3}$N) superimposed on
six color images of different parameter distributions
(T$_{MB}$($^{13}$CO), ratio =
$T_{MB}$($^{13}$CO)/$T_{MB}$(C$^{18}$O), $\tau$($^{18}$CO),
$T_{ex}$($^{13}$CO), N($^{13}$CO)[$\times$10$^{16}$] and 8 $\mu$m).
The contours are at multiples of the 12\%, 13\%, 12\%, 16\%, 14\%,
and 15\% level of each of the HCN, HNC, HCO$^+$, C$_2$H, N$_2$H$^+$,
and HC$_3$N emission peaks, which are 9.573, 7.953, 9.781, 4.791,
10.490, and 2.555 K km s$^{-1}$, respectively. The integrated
velocity range is from $\thicksim$-53.0 to $\thicksim$-43.0 km
s$^{-1}$ for all contours. The black symbols ''$\blacktriangle$'',
''$\blacksquare$'' and ''$\bigstar$'' indicate the positions of the
water maser, the methanol maser, and IRAS 16158-5055, respectively.}
\label{Fig:mopra}
\end{figure*}

\end{appendix}

\end{document}